\documentclass{iaus}
\usepackage{graphicx}

\newcommand{\AD}{\mbox{ATLAS$^{\rm 3D}$}}

\title[The merger origin of ETGs probed by ultra-deep imaging] 
{Investigating  the merger origin of Early-Type Galaxies using  ultra-deep optical images}

\author[Duc et al.]{P.--A.\ Duc,$^1$  J.-C. Cuillandre,$^2$ K.\ Alatalo,$^3$ L.\ Blitz,$^3$ M.\ Bois,$^4$ F.\ Bournaud,$^1$ M.\ Bureau,$^5$  M.\ Cappellari,$^5$  P. C\^ot\'e,$^{6}$ R.\ L.\ Davies,$^5$ T.\ A.\ Davis,$^5$ P.\ T.\ de~Zeeuw,$^{7,8}$ E.\ Emsellem,$^{7,4}$ L.\ Ferrarese,$^6$   E.\ Ferriere,$^1$ S. Gwyn,$^{6}$  S.\ Khochfar,$^{9}$ D.\ Krajnovic,$^7$ H.\ Kuntschner,$^7$ P.-Y.\ Lablanche,$^4$ L. MacArthur,$^{6}$ R.\ M.\ McDermid,$^{10}$ L.\ Michel-Dansac,$^4$ R.\ Morganti,$^{11}$ T. Naab,$^{12}$ T.\ Oosterloo,$^{11}$ M.\ Sarzi,$^{13}$ N.\ Scott,$^5$ P.\ Serra,$^{11}$  A.\ Weijmans$^{14}$ and L.\ M.\ Young$^{15}$}

\affiliation{
$^1$ AIM Paris-Saclay, France;
$^2$ CFHT, USA;
$^3$University of California, Berkeley, USA;
$^4$Observatoire de Lyon, France;
$^5$University of Oxford, UK;
$^6$Herzberg Institute of Astrophysics, Victoria, Canada;
$^7$ESO, Garching, Germany;
$^8$Leiden University, The Netherlands;
$^{9}$MPE, Garching, Germany;
$^{10}$Gemini Observatory, Hilo, USA;
$^{11}$ASTRON, Dwingeloo, The Netherlands;
$^{12}$ MPI, Garching, Germany;
$^{13}$University of Hertfordshire, Hatfield, UK;
$^{14}$Dunlap Institute for Astronomy \& Astrophysics,  University of Toronto, Canada;
$^{15}$New Mexico Tech, Socorro, USA
}

\pubyear{2011}
\volume{277}  
\pagerange{1--1}
\jname{Tracing the Ancestry of Galaxies}
\editors{C. Carignan, F. Combes \& K.C. Freeman, eds.}
\begin{document}

\maketitle

\begin{abstract}
The mass assembly of galaxies  leaves various imprints on their surroundings, such as shells, streams and tidal tails. The  frequency and properties  of these fine structures depend on the mechanism driving the mass assembly: e.g. a monolithic collapse,  rapid cold-gas accretion followed by violent disk instabilities, minor mergers or major dry / wet mergers. Therefore, by studying the outskirts of galaxies, one  can learn about their main formation mechanism. 
I present here our on-going work to characterize the outskirts of Early-Type Galaxies (ETGs), which are  powerful probes at low redshift of the hierarchical mass assembly of galaxies. This work  relies on  ultra--deep optical images obtained at CFHT with the  wide-field of view MegaCam camera of field and cluster ETGs obtained as part of the  \AD\ and NGVS projects. State of the art numerical simulations are used to interpret the data. The  images   reveal a wealth of unknown faint structures at levels as faint as 29 mag arcsec$^{-2}$ in the g-band. Initial results for two galaxies are presented here.

\end{abstract}

\firstsection 
\section{Introduction}

There is a rather large consensus to ascribe to galaxy mergers a prominent role in the formation of  Early-Type Galaxies, especially the most massive ones. For almost 40 years, numerical simulations and  observations have supported this hypothesis. The debate is however still very active  on the details of this merger-based mass assembly: is it primarily due to minor / major, dry / wet mergers? When did the merging events occur, and in which types of environment? 
Study of their internal properties, in particular their kinematics, shows that the properties of ETGs change along the luminosity function, with a general trend for galaxies to transition from slow to fast rotators, while peculiar objects can show distinct features such as Kinematically Decoupled Cores (see the review by E. Emsellem in this volume, and references therein).  Are mergers able to reproduce the full variety of observed properties? Alternatively, it has been proposed that a fraction of  ETGs, especially those with high ellipticity,  might have formed  at high redshift  by secular evolution, i.e. from the interaction, migration and  merging of massive condensations formed in highly turbulent disks, supplied by cold gas flows (\cite[Elmegreen et al., 2008]{Elmegreen08}). 

\begin{figure}[t]
\begin{center}
 \includegraphics[width=\textwidth]{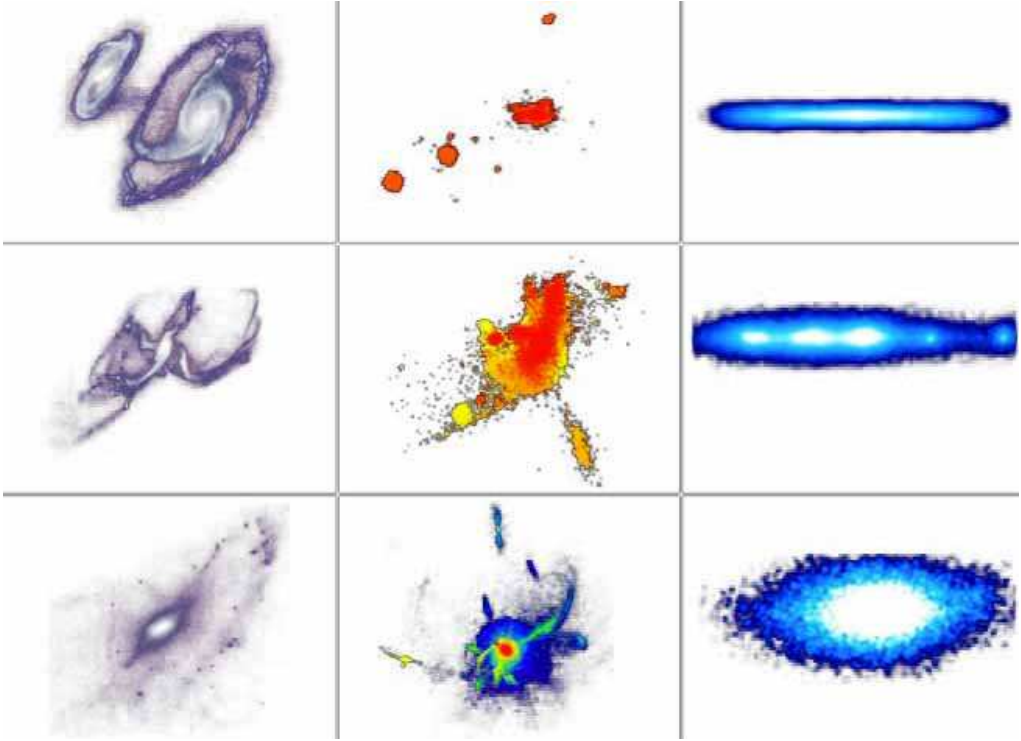} 
 \caption{Predictions from numerical simulations of the frequency, colors and shapes of fine structures for three mass-assembly scenarios. Time sequences are shown for a major merger (left col., \cite[Bournaud et al., 2008]{Bournaud08}), multiple collisions (middle col., from \cite[Martig et al., 2009]{Martig09}) and evolution of an unstable disk (right col., \cite[Elmegreen et al., 2008]{Elmegreen08}). }
   \label{fig:0}
\end{center}
\end{figure}

All these processes produce remnants that have a luminosity profile characterized by a high Sersic index, but with very different morphologies in the outermost regions. While the secular model does not produce any fine structure (Fig.\ref{fig:0}, right), the imprint of (multiple) minor mergers are shells and narrow stellar streams (Fig.\ref{fig:0}, middle) and that of major mergers between spirals are prominent, long, tidal tails (Fig.\ref{fig:0}, left). Thus, by studying the properties of collisional debris, one might  be able to reconstruct the mass assembly of their progenitors. 
This method must, however, address the following issues (1) debris has a very low--surface brightness, usually below 26 mag arcsec$^{-2}$ in the g band  (2)  debris fades with age --  it might  be used as a clock for age-dating merger events  -- or may be dispersed in their environment: the memory of its origin is then lost  (3) galaxies may be formed by hybrid processes. Current detailed numerical simulations done in cosmological context precisely address these issues and are now capable of relating the merging/accretion history of the simulated galaxies with the number, shape, brightness and ages of the fine structures  present at the end of their evolution.

\section{Sample, observations and data processing}
The ultra-deep images presented here were obtained with the large field-of-view MegaCam camera installed on the Canada-France-Hawaii Telescope (CFHT). They were taken as part of two large projects:
\begin{itemize}
\item  $\AD$\footnote{http://www-astro.physics.ox.ac.uk/atlas3d}, a multi-wavelength survey of a complete sample of 260 nearby early-type galaxies within 42 Mpc, complemented by semi-analytical and numerical models (\cite[Cappellari et al., 2011]{Cappellari11}). A sub-sample of approximately 100 ETGs located in the field and in groups were selected for follow-up deep optical imaging in the g' and r' bands.
\item the Next Generation Virgo Cluster Survey (NGVS)\footnote{https://www.astrosci.ca/NGVS/The\_Next\_Generation\_Virgo\_Cluster\_Survey }, which maps the entire Virgo Cluster area in the u*,g',r',i' and z' bands (Ferrarese et al., 2011, in prep.).   Within the 104 square-degrees covered by the  NGVS, there are about 50 ETGs that are part of  the $\AD$ sample.
\end{itemize}
These two surveys reach the same surface brightness limit: about 29 mag arcsec$^{-2}$ in the g-band (about 2.5 mag deeper than the SDSS). Such a sensitivity  could only  be achieved by using specific observing and data-reduction procedures. Basically,  each observation   consisted of multiple  exposures dithered with a very large pattern  (between 7' and 70')   so as to minimize sky background variations. The total exposure time varied between 0.5 and 2 hours depending on the band.  
The data were processed at CFHT with the Elixir Low Surface  Brightness arm of the MegaCam pipeline at CFHT   (Cuillandre et al. 2011, in prep.).

\section{Discovery of new fine structures around ETGs}

\begin{figure}[t]
\begin{center}
 \includegraphics[width=8cm]{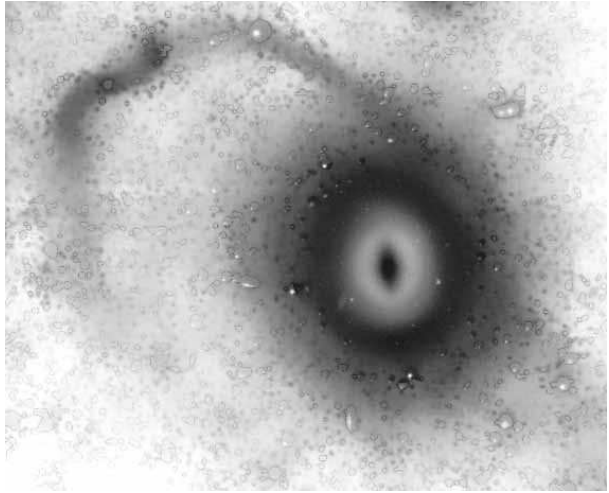} 
 \caption{A satellite galaxy torn away by tidal forces, around an early-type galaxy in the Virgo Cluster. The SDSS image of the galaxy (showing a strong bar) is superimposed on an ultra-deep MegaCam image extracted from the NGVS.  The foreground stars have been subtracted.}
   \label{fig:1}
\end{center}
\end{figure}

\begin{figure}[t]
\begin{center}
 \includegraphics[width=\textwidth]{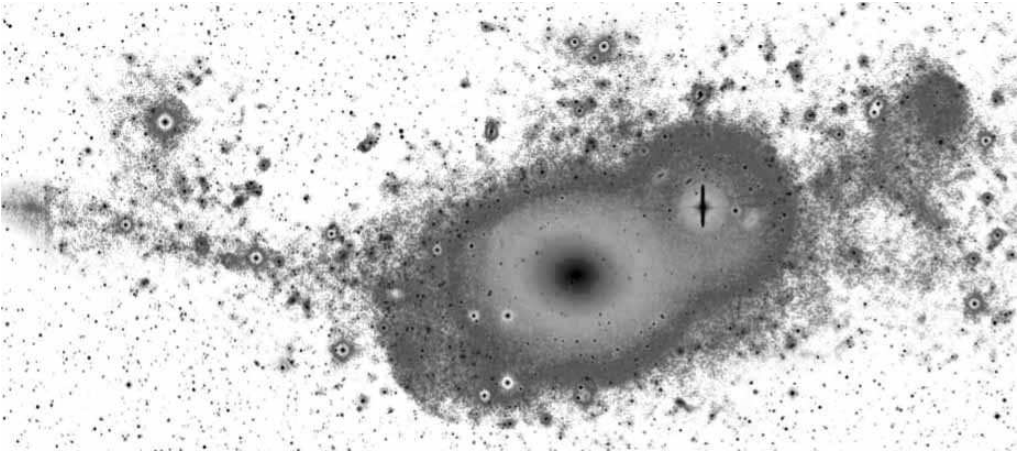} 
 \caption{A massive, round, slow rotator in the \AD\ sample of nearby Early Type Galaxies. The MegaCam image discloses a series of narrow tails and shells that suggest   a rather recent merger. The longest tail has a size of at least 160 kpc. The black stain at the center of the ETG  shows the extent of the galaxy as previously known from the SDSS (Duc et al., 2011) }
   \label{fig:2}
\end{center}
\end{figure}

The unprecedented depth achieved by these surveys  introduces us  to  a new imaging regime, full of faint foreground stars, halos of bright stars, extended, scattered light from galactic cirrus,  distant background galaxies and clusters .... and  previously unknown  stellar streams associated to the early-types galaxies. 
Two examples of newly discovered extended stellar  structures are presented here. Elongated, narrow streams from a disrupted satellite, wrapping around the host galaxy are visible  in Fig.~\ref{fig:1}. This is the first step of a minor merger.  Fig.~\ref{fig:2} discloses prominent, long, tidal tails sticking out of what was so far considered as a massive, relaxed, elliptical, classified as a slow rotator, with the \AD\ definition (\cite[Emsellem et al., 2011]{Emsellem11}). \cite[Duc et al.  (2011)]{Duc11} argue that these  features are likely collisional debris of a relatively recent major merger  that occurred 2 to 5 Gyr ago.  The detection of HI gas along some blue condensations in the tail -- in fact mature Tidal Dwarf Galaxies -- suggests that the progenitors were gas--rich as well. Without the support of  ultra-deep images, such clouds might have been interpreted as signs of on-going gas accretion from cold flows (see review by van Gorkom in this volume for possible examples of such clouds around nearby galaxies).
 The formation of a  massive, round, ETG by a wet merger that occurred below a redshift of 0.5 is at odds with some predictions from semi-analytical models and numerical simulations:   indeed, the formation of such galaxies seems to require a long process involving a high-z major merger followed by multiple minor mergers.\\

These initial results illustrate the potential of ultra-deep optical images to reconstruct the mass assembly of individual galaxies and trace their ancestors.
Furthermore, the \AD\  / NGVS MegaCam  images and follow-up multi-wavelength surveys will provide statistics on  the  number and properties of the fine structures around galaxies, and thus allow to  investigate how  the main mechanism driving the mass assembly may vary as a function of their morphological type and large-scale environment.


\begin{thebibliography}{}


\bibitem[Bournaud \etal\ (2008)]{Bournaud08}
{Bournaud, F., Duc, P.-A. \& Emsellem, E.} 2008,
\textit{MNRAS}, 389, L8
\bibitem[]{Cappellari11}
{Cappellari, M. et al.} 2011,
\textit{MNRAS}, in press (arXiv:1012.1551)
\bibitem[Duc \etal\ (2011)]{Duc11}
{Duc, P-A.,  et al.} 2011,
\textit{MNRAS}, submitted
\bibitem[]{Elmegreen08}
{Elmegreen, B.~G., Bournaud, F. and Elmegreen, D.~M.  } 2008,
\textit{ApJ}, 688, 67
\bibitem[Emsellem \etal\ (2011)]{Emsellem11}
{Emsellem, E..,  et al..} 2011,
\textit{MNRAS}, submitted
\bibitem[]{Martig09}
{Martig, M. et al.} 2009, 
\textit{ApJ}, 707, 250 


\end{thebibliography}
\end{document}